\documentclass[epj]{svjour}
\usepackage{graphics}
\usepackage{doi}

\usepackage{bm}

\usepackage{multirow}
\usepackage{amsmath,amssymb}

\begin{document}
\title{Formation of $\eta'(958)$ Meson Bound States by the $^6\rm{Li}(\gamma,\it{d})$ reaction}


\author{M. Miyatani\inst{1} 
\and N. Ikeno\inst{2} \and H. Nagahiro\inst{1} \and S. Hirenzaki\inst{1}
}                     

%
%
\institute{Department of Physics, Nara Women's University, Nara 630-8506, Japan \and Department of Regional Environment, Tottori University, Tottori 680-8551, Japan}
\date{Received: date / Revised version: date}
%
\abstract{
We have investigated the $^6\rm{Li}(\gamma,\it{d})$ reaction theoretically for the formation of the $\eta'(958)$ mesic nucleus close to the recoilless kinematics. We have developed the theoretical formula and reported the quantitative results of the formation spectra for various cases in this article. We have found that the formation cross sections are reduced by the effects of the fragile deuteron form factor.
} 
\maketitle

\section{Introduction}
The study of the symmetry breaking pattern of QCD and its (partial) restoration in extreme conditions such as high density $\rho$ and/or temperature $T$ is one of the most interesting subjects in the contemporary hadron nuclear physics \cite{Hatsuda1994}. Within many researches in this field, the study of the meson--nucleus bound states has certain advantages and enables us to perform the selective observation of the meson properties with fixed quantum numbers of the bound states. They also enable us to obtain the quasi-static information without contaminations by the dynamical time evolution of the system \cite{Yamazaki2012}. Thus, the study of the meson--nucleus bound systems is one of the best methods to investigate the meson properties at finite density $\rho \lesssim 0.17\ \rm{{fm}^{-3}}$ with $T=0$.

From the successful observation of the deeply bound pionic atoms \cite{Hirenzaki1991,Gilg,Itahashi2000}, one has recognized well that it is critically important to find out the way how we can produce and observe new meson--nucleus bound states in laboratory. It is essential to develop and widen this field to new frontiers such as the study of the heavier meson properties in nucleus. So far, one-nucleon transfer reactions have been mainly considered and successfully used to produce the nuclear bound states of a meson in the recoilless kinematics \cite{Yamazaki2012,Hirenzaki1991}. It has been known that the formation of the bound states in the recoilless kinematics is most efficient and important for the experimental observation. 

However, the recoilless condition can be satisfied only for lighter mesons than nucleon, in the one-nucleon transfer reactions. To overcome this difficulty, Ikeno \it et al. \rm studied the two nucleon pick-up reactions on the $^6\rm{Li}$ target for the formation of heavy meson bound states with the $\alpha$ particle \cite{Ikeno_gamma_d}. We develop this study further in this article and improve the theoretical method in the following four points;  \\
(i) the distortion effects are taken into account for emitted deuteron,  \\
(ii) the elementary cross section is evaluated phenomenologically and the absolute value of the formation cross section is obtained,  \\
(iii) the realistic $\alpha$ density distribution is used to calculate the meson--$\alpha$ bound states,  \\
(iv) the correction of the center-of-mass motion in the calculation of the effective numbers is considered.  \\
We formulate the improved theoretical model and show the quantitative numerical results. 

As for the actual meson which we investigate nuclear bound system formation in this article, we consider the $\eta'(958)$ ($\eta'$) meson. The $\eta'$ meson is a heavier meson than nucleon and is believed to get so heavy because of the $\rm{U}_{\it{A}}(1)$ anomaly. Recently, the structure and formation of the bound states of $\eta'$ in nucleus have been studied in theoretical and experimental points of view \cite{Nagahiro2005,Nagahiro2013,Itahashi2012} in the contexts of the strong symmetry and the $\eta'$ property changes, especially its mass shift in nucleus. In the present exploratory level, the study of the $\eta'$ bound states formation is important and necessary. In this article, we consider the formation of the $\eta'$ bound states in the $\alpha$ particle in the $^6\rm{Li}(\gamma,\it{d})$ reaction and report the calculated results of the formation cross section. It should also be noted that the $\eta$ meson--nucleus systems have been studied for a long time both theoretically and experimentally \cite{Kelkar2013,Krusche2014,Machner2015,Haider2015}.

This paper is organized as follows. In section \ref{sec:Effective number formalism}, we give the formulation to get the formation rate of the $\eta'$ bound states in the effective number approach for the $^6\rm{Li}(\gamma,\it{d})$ reaction. Then, in section \ref{sec:Numerical results} we show the numerical results of the $\eta'$ mesic nucleus formation cross section, where we report the results of the elementary cross section, the structure of the $\eta'$--$\alpha$ bound systems, and the calculated formation spectra. We also give some discussions on the effects of the possible shrinkage of the quasi-deuteron in the $^6\rm{Li}$ target. We devote section \ref{sec:Conclusions} to summarize this work.

\section{Effective number formalism for the quasi-deuteron in nucleus} \label{sec:Effective number formalism}
The theoretical calculation of the two-nucleon transfer reactions is rather difficult in general. Thus, we have adopted same theoretical formula as in Ref. \cite{Ikeno_gamma_d} and considered the $^6\rm{Li}$ target, which is expected to have large probability of the $\alpha+d$ component in its ground state. The probability is reported to be 0.73 in Ref. \cite{Ent_0.73}.

We apply the effective number approach to evaluate the formation rate of the $\eta'$--nucleus bound system in the $(\gamma,\it{d})$ reaction as in Ref. \cite{Ikeno_gamma_d}. We evaluate the formation cross section in the laboratory frame as a function of the energy $E_d$ and the solid angle $\Omega_d$ of the emitted deuteron by applying the effective number approach. The cross section can be written as,

\begin{eqnarray}
\frac{d^2\sigma}{dE_dd\Omega_d}=\left(\frac{d\sigma}{d\Omega_d}\right)^{\rm{ele}}_{\rm{Lab}}\sum_f N_{\rm{eff}}\frac{\Gamma_{\eta'}}{2\pi}\frac{1}{\Delta E^2+\Gamma_{\eta'}^2/4}, \label{eq:CS}
\end{eqnarray}

\noindent where $\left(d\sigma/d\Omega_d\right)^{\rm{ele}}_{\rm{Lab}}$ is the elementary cross section of the $\eta'$ meson photo production from deuteron, $\gamma+d \rightarrow d+\eta'$ reaction. In this formalism, we need the value of the elementary cross section at angles with the largest momentum transfer to the $\it{deuteron}$ in the elementary process since the deuteron must be emitted to forward angles in the laboratory frame with similar momentum with the incident photon in the recoilless kinematics of $^6\rm{Li}(\gamma,\it{d})$ reaction. The final states with $\alpha$ and $\eta'$ indicated by $\it f$ are summed up to evaluate the $\gamma+^6\rm{Li} \rightarrow \it{d}+(\eta'$--$\alpha)$ cross section. We have summed up the contributions of all the bound $\eta'$--$\alpha$ states with the width $\Gamma_{\eta'}$ listed in Table \ref{table:BE-gamma}. The $\eta'$--$\alpha$ scattering states are not included in our calculations, which contribute to the spectrum above the threshold. The energy transfer $\Delta E$ of the reaction in the laboratory frame is defined as, 

\begin{eqnarray}
\Delta E=T_d-p_\gamma+S_d-B_{\eta'}+m_{\eta'}, \label{eq:delE}
\end{eqnarray}

\noindent where $T_d$ is the emitted deuteron kinetic energy, $p_\gamma$ the incident photon momentum, and $m_{\eta'}$ the $\eta'$ meson mass. The $\eta'$ meson binding energy $B_{\eta'}$ is determined for each bound level of the $\eta'$ meson and the deuteron separation energy $S_d$ is evaluated to be $S_d=1.47\ \rm{MeV}$ by calculating the mass gap as $S_d=(M_\alpha+M_d)-M_{^6\rm{Li}}$. Here, we neglect the recoil energy of the daughter nucleus in this expression of the reaction kinematics since we mainly consider the kinematics close to the recoilless condition.

The effective number in Eq. (\ref{eq:CS}) of the $^6\rm{Li}(\gamma,\it{d})\alpha\otimes\eta'$ reaction can be written as,

\begin{eqnarray}
N_{\rm{eff}}=\sum_{JM}\left|\int\chi_{d'}^\ast(\bm{r})\left[\phi_{l_{\eta'}}^\ast(\bm{r})\otimes\psi_{l_d}(\bm{r})\right]_
{JM}\chi_\gamma(\bm{r})d\bm{r}\right|^2,  \nonumber \\
\label{eq:Neff}
\end{eqnarray}

\noindent where $\phi_{l_{\eta'}}(\bm{r})$ and $\psi_{l_d}(\bm{r})$ are the wave functions of the $\eta'$ meson and the deuteron bound to $\alpha$. $\chi_\gamma(\bm{r})$ and $\chi_{d'}(\bm{r})$ are the incident photon and the emitted deuteron wave functions in the scattering states, respectively. The effective number is considered to evaluate the effective nucleon number participating the reaction and to provide an expression of the nucleon response function.  

\begin{figure}[htbp]
\begin{center}
\resizebox{0.45\textwidth}{!}{%
\includegraphics{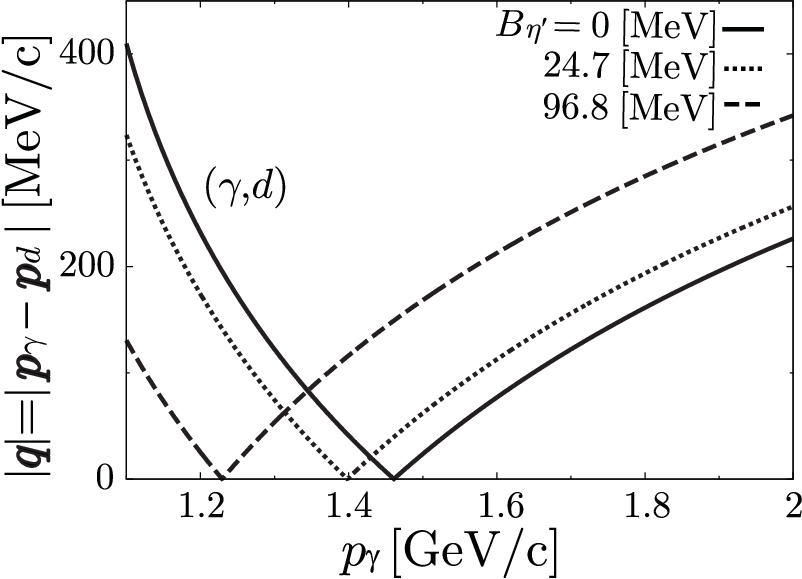}
}
\end{center}
\caption{Momentum transfer $\bm{q}$ of the forward $^6\rm{Li}(\gamma,\it{d})$ reaction for the $\eta'$ meson bound state formation as functions of the incident photon momentum $p_\gamma$ for three values of the $\eta'$ meson binding energy $B_{\eta'}$. The recoil energy of the daughter nucleus is neglected. The deuteron separation energy $S_d$ is fixed to be $S_d=1.47\ \rm{MeV}$.}
\label{fig:q}
\end{figure}

The momentum transfer $\bm{q}$ of the reaction is defined as, 

\begin{eqnarray}
\bm{q}=\bm{p}_{\gamma} - \bm{p}_d,
\end{eqnarray}

\noindent and it is shown in Fig. \ref{fig:q} for the formation of an $\eta'$ meson bound state in the cases of $B_{\eta'}=0, 24.7, 96.8\ \rm{MeV}$. $\bm{q}$ is calculated by considering the kinematics with $\Delta E=0$ in Eq. (\ref{eq:delE}) in the laboratory frame. The momentum transfer $\bm{q}$ with $B_{\eta'}=0$ corresponds to the threshold $\eta'$ production, and $\bm{q}$ with $B_{\eta'}=24.7\ \rm{MeV}$ and $96.8\ \rm{MeV}$ correspond to the formation of the $(\eta'$--$\alpha)$ bound states obtained in section \ref{sec:Numerical results}. From the figure, we find that the $\eta'$ meson at the threshold $(B_{\eta'}=0)$ is produced in the recoilless kinematics by the $(\gamma,\it{d})$ reaction with $E_\gamma\sim1.46\ \rm{GeV}$, while bound $\eta'$ meson with binding energies $B_{\eta'}=24.7$ and $96.8\ \rm{MeV}$ can be produced in the recoilless kinematics by the incident photon with $E_\gamma\sim1.40$ and $1.23\ \rm{GeV}$, respectively. Thus, it is confirmed that the $\eta'$ meson can be produced in the recoilless kinematics in the $(\gamma,\it{d})$ reaction by choosing the appropriate photon energy as pointed out in Ref. \cite{Ikeno_gamma_d}.

\begin{figure}[htbp]
\begin{center}
\resizebox{0.35\textwidth}{!}{%
\includegraphics{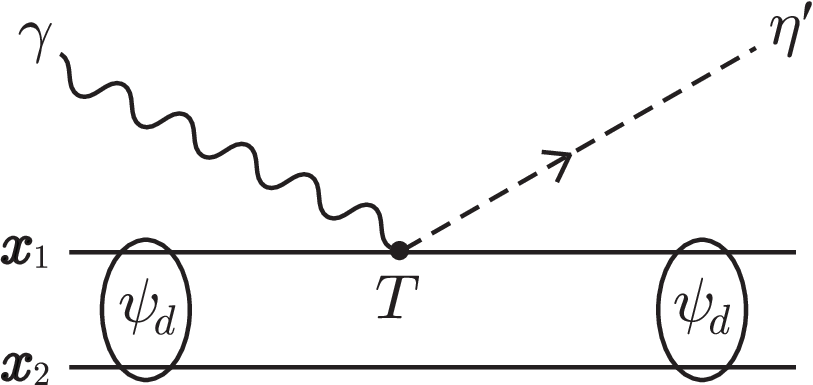}
}
\end{center}
\caption{Schematic figure of the elementary process of the coherent $\eta'$ meson photo-production from the deuteron target considered in this article. $\psi_d$ and $T$ indicate the anti-symmetrized deuteron wave function and the transition amplitude of the $\gamma+N \rightarrow \eta'+N$ process.}
\label{fig:elementary_process}
\end{figure}

The elementary cross section $\left(d\sigma/d\Omega_d\right)^{\rm{ele}}_{\rm{Lab}}$ in Eq. (\ref{eq:CS}) is the cross section of the $\gamma+d \rightarrow d+\eta'$ reaction. We show the schematic figure of this reaction in Fig. \ref{fig:elementary_process}. Since the data of the differential cross section in this kinematics are not available unfortunately, we consider a simple model as indicated in Fig. \ref{fig:elementary_process} to evaluate the elementary cross section. The $S$--matrix can be written as,

\begin{eqnarray}
S &=& \int \sqrt{\frac{M_d}{E_{d'}}} \frac 1{\sqrt{V}} \mathrm{exp}\left[-i\bm{p}_{d'} \cdot \left( \frac{\bm{x}_1+\bm{x}_2}{2}\right)\right]\psi_d^\ast \left(\bm{x}_1-\bm{x}_2\right) \nonumber \\
&\times& \frac 1{\sqrt{2\omega_{\eta'}}} \frac 1{\sqrt{V}} \mathrm{exp} \left[ -i\bm{p}_{\eta'} \cdot \bm{x}_1 \right] iT  \nonumber \\ 
&\times& \sqrt{\frac{M_d}{E_d}} \frac 1{\sqrt{V}} \mathrm{exp} \left[ i\bm{p}_d \cdot \left( \frac{\bm{x}_1+\bm{x}_2}{2} \right) \right] \psi_d \left(\bm{x}_1-\bm{x}_2\right) \nonumber \\
&\times& \frac 1{\sqrt{2E_\gamma}} \frac 1{\sqrt{V}} \mathrm{exp} \left[ i\bm{p}_\gamma \cdot \bm{x}_1 \right]  \nonumber \\
&\times& (2\pi) \delta\left( E_d + E_\gamma - E_{d'} - \omega_{\eta'} \right) d\bm{x}_1 d\bm{x}_2,  \label{eq:S}
\end{eqnarray}

\noindent where the coordinates of the initial nucleons are written as $\bm{x}_1$ and $\bm{x}_2$, $M_d$ indicates the deuteron mass, $E_d$ and $E_{d'}$ the deuteron energies in the initial and final states, and $\omega_{\eta'}$ the emitted $\eta'$ meson energy. $\bm{p}_d, \bm{p}_{d'}, \bm{p}_{\eta'},$ and $\bm{p}_\gamma$ indicates the momentum of the each particle. $\psi_d$ and $T$ indicate the anti-symmetrized deuteron wave function and the transition amplitude of the $\gamma+N \rightarrow \eta'+N$ process. Since the deuteron wave function has the largest spatial dimensions in this system and the deuteron form factor is expected to determine the angular dependence of the cross section, we assume the transition amplitude $T$ to be expressed by a coupling constant $c$ and put $T=c$, which will be determined by the available experimental data.

Using the $S$--matrix, we can obtain the elementary cross section by performing the phase space integration,

\begin{eqnarray}
d\sigma=\frac {V}{v_{\mathrm{rel}}} \frac {\left| S \right|^2}{T} \frac V{(2\pi)^3} d\bm{p}_{d'} \frac V{(2\pi)^3} d\bm{p}_{\eta'},  \label{eq:dsigma}
\end{eqnarray}

\noindent where $v_{\mathrm{rel}}$ indicates the relative velocity between the initial photon and deuteron. The cross section in the center-of-mass frame of the $\gamma+d \rightarrow d+\eta'$ reaction can be finally written as,

\begin{eqnarray}
\left(\frac{d\sigma}{d\Omega_d}\right)^{\rm{ele}}_{\rm{CM}}=\frac{|c|^2}{8\pi^2}
\frac{M_d^2}{\lambda^{1/2}(s,M_d^2,0)}\frac{p_{d'}}{E_{d'}+\omega_{\eta'}}|F_d(\bm{q})|^2, \label{eq:eleCS}
\end{eqnarray}

\noindent where $F_d$ is the deuteron form factor. It should be noted that the momentum transfer $\bm{q}$ in $F_d$ should be evaluated in the initial deuteron rest frame, though other kinematical variables in Eq. (\ref{eq:eleCS}) are defined in the center-of-mass frame of the reaction. $\lambda$ indicates the K\"{a}llen function defined as,

\begin{eqnarray}
\lambda(x,y,z)=x^2+y^2+z^2-2xy-2yz-2zx.
\end{eqnarray}

\noindent In section \ref{sec:Numerical results}, the applicability of this simple formula is checked using the $\gamma+d \rightarrow d+\eta$ data, and the coupling constant $c$ of the contact interaction in Eq. (\ref{eq:eleCS}) is determined by the data of the $\eta'$ production total cross section in Ref. \cite{totCS_etaprime}.
 
 The deuteron form factor $F_d$ is defined as, 

\begin{eqnarray}
\left|F_d(\bm{q})\right|^2=\frac13 \sum_{\rm{M'M}}\left|F^{\rm{M'M}}_d(\bm{q})\right|^2, 
\end{eqnarray}

\noindent where $\rm{M}$ and $\rm{M'}$ indicate the magnetic quantum number of the deuteron total spin in the initial and final states, respectively. $F^{\rm{M'M}}_d$ can be calculated as,

\begin{eqnarray}
F^{\rm{M'M}}_d(\bm{q})=\int\psi_d^{\ast\rm{M'}}(\bm{r})e^{i\bm{q} \cdot \bm{r}/2}\psi^{\rm{M}}_d(\bm{r})d\bm{r},
\end{eqnarray}

\noindent using the standard deuteron wave function $\psi^{\rm{M}}_d$, which is written as the sum of the S- and D-wave parts as,

\begin{eqnarray}
\psi^{\rm{M}}_d(\bm{r})=\psi^{\rm{M}}_S(\bm{r})+\psi^{\rm{M}}_D(\bm{r}).
\end{eqnarray}

\noindent The S- and D-wave part wave functions are written with two nucleon spin wave function $\chi_s^{\rm{M}_{\it{s}}}$ as,

\begin{eqnarray}
\psi^1_S(r)&=&\frac {u(r)}{r}\frac 1{\sqrt{4\pi}}\chi^1_1, \\
\psi^1_D(r)&=&\frac {w(r)}{r}\left(\sqrt{\frac 35}Y^2_2\chi^{-1}_1 \right. \nonumber \\
              &-& \left. \sqrt{\frac 3{10}}Y^1_2\chi^{0}_1+\sqrt{\frac 1{10}}Y^0_2\chi^1_1\right), 
\end{eqnarray}

\noindent for the $\rm{M}=1$ state, for example. The expressions of all $F_d^{\rm{M'M}}$ can be found in Ref. \cite{Fellinger} as $S_{m',m}$ in their notation. After some manipulations, the form factor $\left|F_d\right|^2$ can be written as,

\begin{eqnarray}
\left|F_d(\bm{q})\right|^2=\left|S_0\left(\frac12q\right)\right|^2+\left|S_2\left(\frac12q\right)\right|^2, \label{eq:F_S0,2}
\end{eqnarray}

\noindent where $S_0$ and $S_2$ are defined as, 

\begin{eqnarray}
S_0(p)=\int^{\infty}_0\left[u^2(r)+w^2(r)\right]j_0(pr)dr, 
\end{eqnarray}

\noindent and

\begin{eqnarray}
S_2(p)=\int^{\infty}_0\left[2u(r)w(r)-w^2(r)/{\sqrt{2}}\right]j_2(pr)dr,
\end{eqnarray}

\noindent with $j_0$ and $j_2$ the spherical Bessel functions. As for the deuteron radial wave functions $u(r)/r$ and $w(r)/r$, we use those obtained by the Bonn potential \cite{Bonn}. The momentum transfer $\bm{q}$ to the deuteron in Eq. (\ref{eq:F_S0,2}) is evaluated in the initial deuteron rest frame in the elementary process shown in Fig. \ref{fig:elementary_process}.

We take into account the distortion effects to the deuteron wave function by the Eikonal approximation and $\chi_{d'}^\ast(\bm{r})\chi_\gamma(\bm{r})$ in Eq. (\ref{eq:Neff}) is written as, 

\begin{eqnarray}
\chi_{d'}^\ast(\bm{r})\chi_\gamma(\bm{r})=e^{i\bm{q} \cdot \bm{r}}D(\bm{b},z), \label{eq:scatteringwf}
\end{eqnarray}

\noindent where $\it{D}$ is the distortion factor defined as,

\begin{eqnarray}
D(\bm{b},z)=\mathrm{exp} \left[- \frac{\sigma_{dN}}{2} \int_z^{+\infty} \rho_{\alpha}(\bm{b},z')dz' \right],
\end{eqnarray}

\noindent with the impact parameter $\bm{b}$ and the beam direction coordinate  $z$. The distortion cross section $\sigma_{dN}$ is fixed to be $\sigma_{dN}=60\ \rm{mb}$ \cite{PDG}. $\rho_\alpha$ indicates the density distribution of the daughter nucleus. The emitted deuteron could also be distorted by the interaction with bound $\eta'$ meson. We do not consider this effect here because of the lack of the information on $\eta'$ interaction in this exploratory level. To take into account the correction due to the center-of-mass motion in the calculation of $N_{\rm{eff}}$, we have scaled the coordinate $\bm{r}$ in Eq. (\ref{eq:scatteringwf}) as,

\begin{eqnarray}
\bm{r} \rightarrow \frac{M_\alpha}{m_{\eta'}+M_\alpha}\, \bm{r},
\end{eqnarray}

\noindent with the masses of the $\eta'$ meson $m_{\eta'}$ and the $\alpha$ particle $M_\alpha$ according to the prescription adopted in Ref. \cite{Koike}.

The bound $\eta'$ meson wave functions in the final state are calculated by solving the Klein-Gordon equation with an optical potential $U_{\eta'}(r)$ written as,

\begin{eqnarray}
U_{\eta'}(r)=(V_0+iW_0)\frac{\rho_\alpha(r)}{\rho_0},  \label{eq:potential-eta'}
\end{eqnarray}

\noindent where $V_0$ and $W_0$ are the parameters which determine the real and imaginary potential strength at the normal nuclear density
$\rho_0=0.17\ \rm{{fm}^{-3}}$, respectively. 

The strength of the $\eta'$--nucleus potential is still controversial and has not been determined well. In Refs. \cite{Nagahiro2005,Nagahiro2006}, where the formation of the $\eta'$ mesic nucleus was considered for the first time, the strength of the $\eta'$--nucleus potential was evaluated by the NJL model to be around $-150\ \rm{MeV}$. Actually the recent evaluation based on the chiral symmetry restoration \cite{Jido2012} also indicates the strong attractive and less absorptive potential. Another calculation based on the chiral unitary model \cite{Nagahiro2012} reveals the sensitivity of the potential to the coupling strength of the singlet $\eta$ to the octet baryons. On the other hand, the latest experimental data indicate the small $\eta' N$ scattering length \cite{Czerwinski2014} and the shallow $\eta'$--nucleus potential \cite{Nanova2012,Nanova2013}. Theoretical evaluation in Ref. \cite{Bass2014} also indicates weak attractive potential. Therefore, we accept large uncertainties of the $\eta'$--nuclear optical potential strength and consider wide varieties of potential strength as introduced in section \ref{subsec:structure}, which almost cover the potential strength mentioned above.  

As for the density distribution of the $\alpha$ particle $\rho_\alpha$, we have used the theoretical results obtained by the Gaussian expansion method \cite{Hiyama_private,Hiyama}, which is known as one of the most accurate method for the studies of the few-body systems. 

The wave function for the relative motion of deuteron and $\alpha$ in the $^6\rm{Li}$ target $\psi_{l_d}$ is determined to reproduce the momentum distribution reported in Ref. \cite{Ent_0.73} as in Ref. \cite{Ikeno_gamma_d}, and is obtained by solving the Schr\"{o}dinger equation with the Woods-Saxon type potential,

\begin{eqnarray}
U(r)=\frac{U_0}{1+\exp((r-R)/a)},
\end{eqnarray}

\noindent where the parameters are fixed to be $R=2.0\ \rm{fm}$, $a=0.5\ \rm{fm}$, and $U_0=-75\ \rm{MeV}$ as in Ref. \cite{Ikeno_gamma_d}.

Finally, we should mention here that the expression of Eq. (\ref{eq:CS}) for the bound state formation cross section is based on the factorization assumption of the elementary meson production cross section and the effective numbers which describe the nuclear response. Since the elementary cross section is determined to reproduce the on-shell $\eta'$ production data as described below, the off-shell effects of the elementary process have not been taken into account in the present formalism, which could affect the numerical results reported in this article \cite{Kelkar2013}.

\section{Numerical results} \label{sec:Numerical results}
\subsection{Elementary Cross Section}
\begin{figure}[htbp]
\begin{center} 
\resizebox{0.45\textwidth}{!}{%
\includegraphics{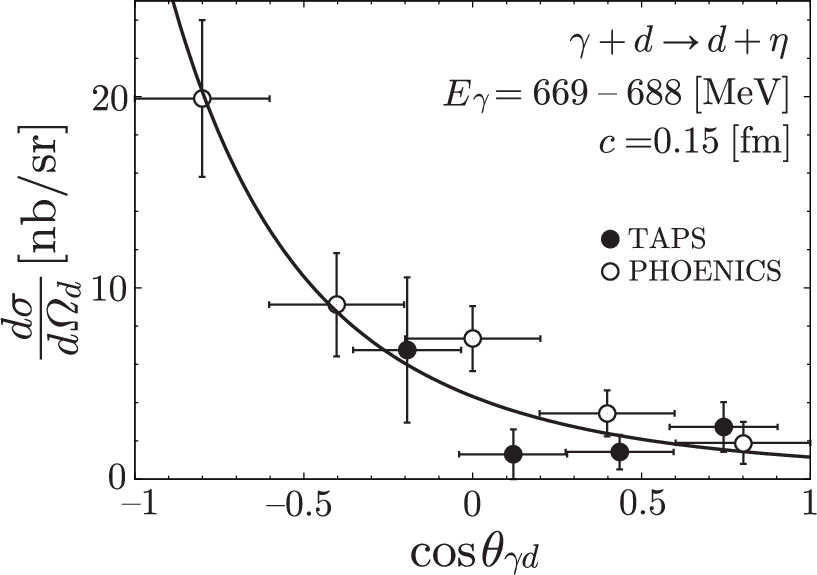}
}
\end{center}
\caption{Angular distribution of the coherent $\eta$ production with the photon energy $E_\gamma=669-688\ \rm{MeV}$ from the deuteron target in the center-of-mass frame. Experimental data are taken from \cite{eleCS_eta} (TAPS) and \cite{eleCS_eta_data} (PHOENICS). The solid line is the calculated result at $E_\gamma=678.5\ \rm{MeV}$ by the present model shown in Eq. (\ref{eq:eleCS}) for $\eta$ production with $c=0.15\ \rm{fm}$.}
\label{fig:eleCScm-eta}
\end{figure}

\begin{figure}[htbp]
\begin{center}
\resizebox{0.45\textwidth}{!}{%
\includegraphics{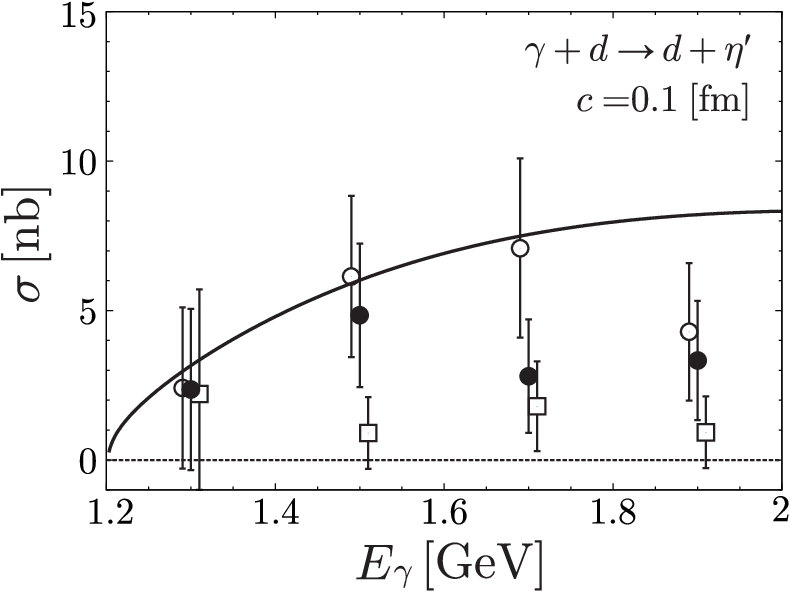}
}
\end{center}
\caption{Total cross section for the coherent $\eta'$ production from the deuteron target as a function of the incident photon energy. Experimental data are taken from Fig. 14 in Ref. \cite{totCS_etaprime}. Different symbols correspond to different analyses in Ref. \cite{totCS_etaprime}. The solid line is the calculated result by the present model shown in Eq. (\ref{eq:eleCS}) with $c=0.1\ \rm{fm}$.}
\label{fig:totCS-etaprime}
\end{figure}

In this subsection, we show the numerical results of the elementary cross sections and study the validity of our simple formula shown in Eq. (\ref{eq:eleCS}). Since we have no experimental data of the angular distribution of the $\gamma+d \rightarrow d+\eta'$ reaction, we apply our model to the $\eta$ production and compare the results to the data of the $\gamma+d \rightarrow d+\eta$ reaction in Fig. \ref{fig:eleCScm-eta} to study the applicability of the model. As we can see in Fig. \ref{fig:eleCScm-eta}, the angular dependence of the coherent $\eta$ production from deuteron target is well reproduced with the coupling strength parameter $c=0.15\ \rm{fm}$. We have also checked our numerical results for other cases with different photon energies reported in Ref. \cite{eleCS_eta,eleCS_eta_data}, where we can also find the experimental angular distribution of the $\gamma+d \rightarrow d+\eta$ reaction with photon energy intervals $E_\gamma=629-649$, $649-669$, and $688-716\ \rm{MeV}$. We have found the good reproduction again for other energies with $c=0.13-0.15\ \rm{fm}$. Then, we have compared our results to the data of the total cross section of $\gamma+d \rightarrow d+\eta'$ reaction as shown in Fig. \ref{fig:totCS-etaprime}. We can see from the figure that the energy dependence of the total cross section is reasonably reproduced around $E_\gamma=1.2-1.5\ \rm{GeV}$ with $c=0.1\ \rm{fm}$. Thus, we have used the model shown in Eq. (\ref{eq:eleCS}) to evaluate the
angular dependence of the elementary cross section which is necessary for the effective number approach.

\begin{figure}[htbp]
\begin{center}
\resizebox{0.45\textwidth}{!}{%
\includegraphics{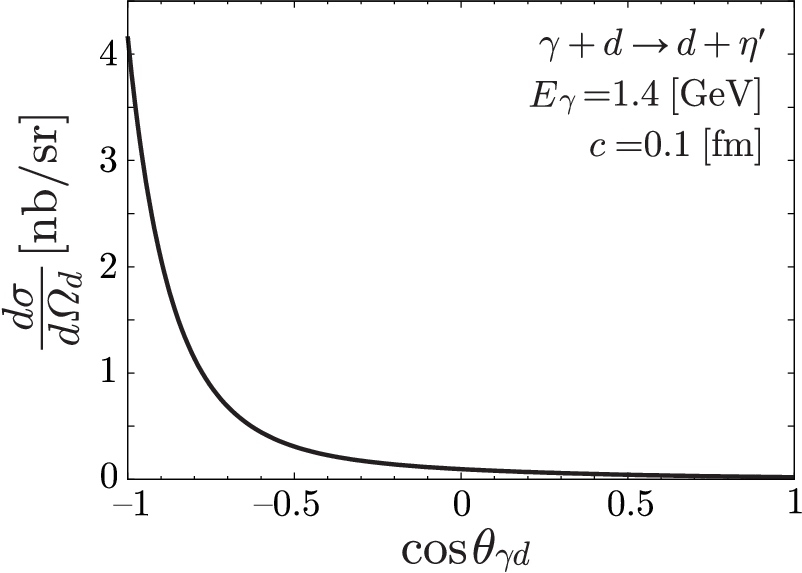}
}
\end{center}
\caption{Angular distribution of the coherent $\eta'$ production with the photon energy $E_\gamma=1.4\ \rm{GeV}$ for the deuteron target in the center-of-mass frame. The solid line is the calculated result by the present model shown in Eq. (\ref{eq:eleCS}) with $c=0.1\ \rm{fm}$.}
\label{fig:eleCScm-etaprime}
\end{figure}

We show the calculated cross section of the $\gamma+d \rightarrow d+\eta'$ reaction at $E_\gamma=1.4\ \rm{GeV}$ as a function of the scattering angle $\theta_{\gamma d}$ in the center-of-mass frame in Fig. \ref{fig:eleCScm-etaprime}. Since the scattering angle $\theta_{\gamma d}$ is defined as that between the incident photon and the emitted deuteron, the momentum transfer takes the maximum value and the cross section the minimum value at $\theta_{\gamma d}=0^\circ$. The value of the elementary cross section at $\theta_{\gamma d}=0^\circ$ which is used in the effective number approach is $0.019\ \rm{nb/sr}$ in the center-of-mass frame and $0.208\ \rm{nb/sr}$ in the laboratory frame at $E_\gamma=1.4\ \rm{GeV}$.  

\begin{figure}[htbp]
\begin{center}
\resizebox{0.45\textwidth}{!}{%
\includegraphics{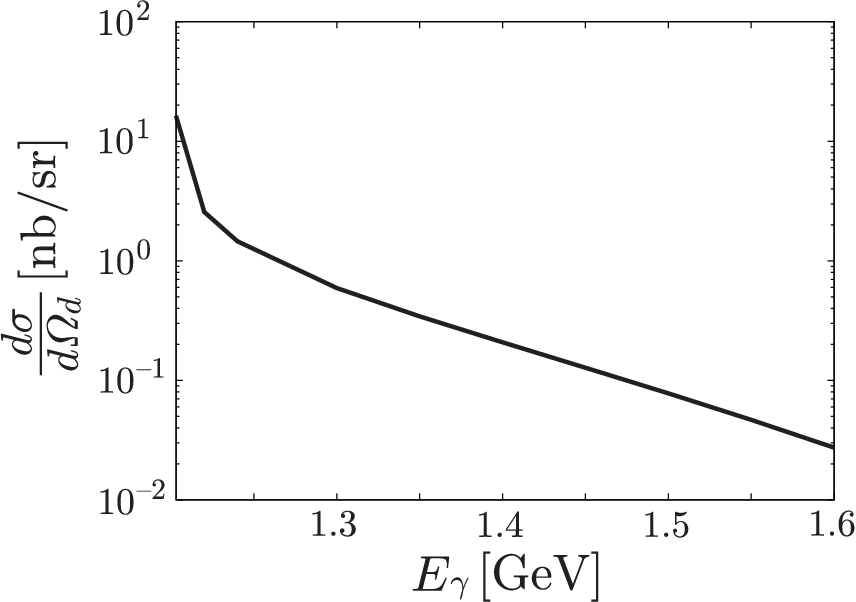}
}
\end{center}
\caption{Incident photon energy dependence of the differential cross section of the $\gamma+d \rightarrow d+\eta'$ reaction at $\theta_{\gamma d}=0^\circ$ in the laboratory frame with $c=0.1\ \rm{fm}$.}
\label{fig:etaprime-eleCS-0}
\end{figure}

Then, we show in Fig. \ref{fig:etaprime-eleCS-0} the incident photon energy dependence of the elementary cross section at $\theta_{\gamma d}=0^\circ$. We can see from the figure that the elementary cross section at $\theta_{\gamma d}=0^\circ$ becomes significantly smaller for higher photon energies because of the fragile deuteron form factor. For the higher photon energies, the total cross section becomes larger as already shown in Fig. \ref{fig:totCS-etaprime} because of the larger phase volume. However, at the same time, the momentum transfer at $\theta_{\gamma d}=0^\circ$ becomes larger and the cross section has more backward peak structure for higher photon energies. Consequently, the differential cross section at $\theta_{\gamma d}=0^\circ$ behaves as shown in Fig. \ref{fig:etaprime-eleCS-0} as a function of the incident photon energy. We use the cross section shown in this figure as the elementary cross section in the effective number approach.

\subsection{Structure of the $\eta'$--$\alpha$ bound states} \label{subsec:structure}
\begin{figure}[htbp]
\begin{center}
\resizebox{0.45\textwidth}{!}{%
\includegraphics{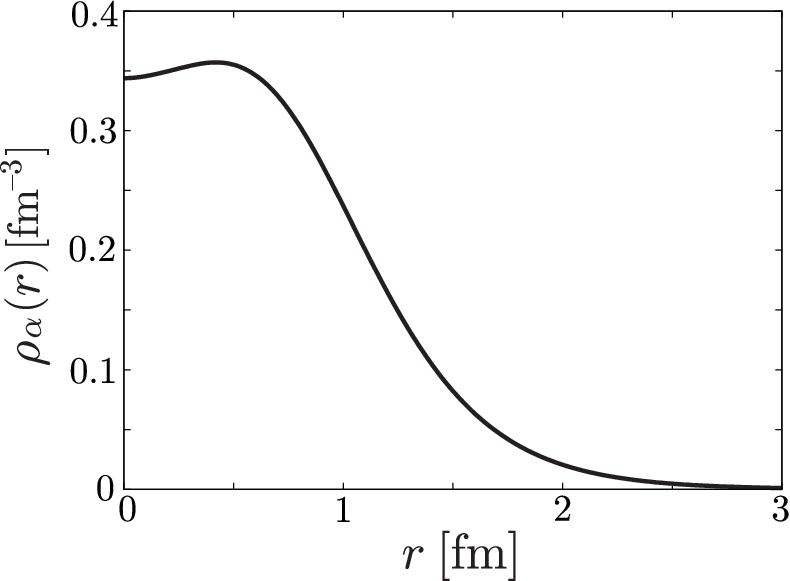}
}
\end{center}
\caption{Realistic $\alpha$ density distribution calculated by the Gaussian expansion method \cite{Hiyama_private,Hiyama}.}
\label{fig:rho-alpha}
\end{figure}

The structure of the $\eta'$--$\alpha$ bound states are calculated by solving the Klein-Gordon equation with the optical potential given in Eq. (\ref{eq:potential-eta'}). We have used a realistic density distribution of the $\alpha$ particle \cite{Hiyama_private,Hiyama}, which is shown in Fig. \ref{fig:rho-alpha} as a function of the radial coordinate $r$. We should mention here that the central density of $\alpha$ is as huge as $\rho_\alpha (0)=0.34\ \rm{{fm}^{-3}}$, as well-known, and is twice as the normal nuclear density $\rho_0=0.17\ \rm{{fm}^{-3}}$.

\begin{figure*}[htbp]
\begin{center}
\resizebox{0.75\textwidth}{!}{%
\includegraphics{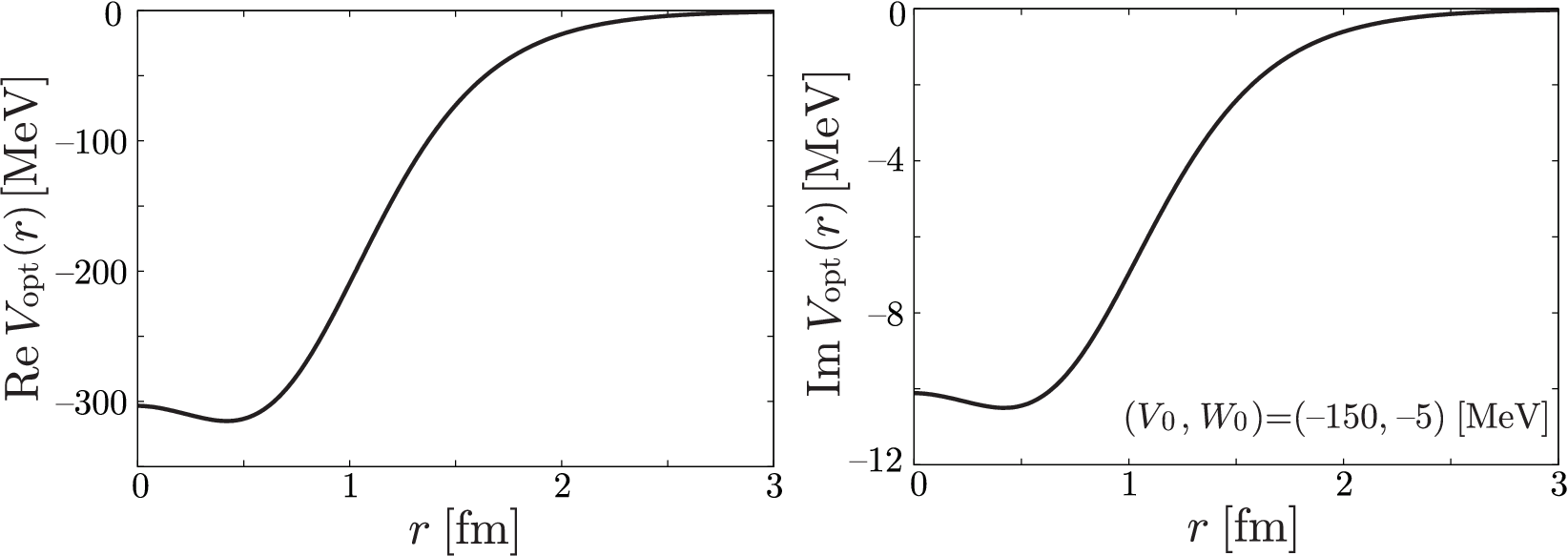}
}
\end{center}
\caption{Profile of the $\eta'$--$\alpha$ optical potential is shown as functions of the radial coordinate $r$ for the real part (left) and the imaginary part (right) for the potential parameters $(V_0, W_0)=(-150, -5)\ \rm{MeV}$.}
\label{fig:V0-3}
\end{figure*}

As for the potential parameters, $V_0$ and $W_0$, we consider the wide varieties of the potential strength as discussed in section \ref{sec:Effective number formalism} and use the same sets of $V_0$ and $W_0$ as in Ref. \cite{Nagahiro2013}. We show the potential profile in Fig. \ref{fig:V0-3} for $(V_0, W_0)=(-150, -5)\ \rm{MeV}$ case, as an example. As we can see in the figure, we should note that the central strength of the potential is large and is about $-300\ \rm{MeV}$ for real part and $-10\ \rm{MeV}$ for imaginary part because of the high central density in $\alpha$.

\begin{figure}[htbp]
\begin{center}
\resizebox{0.4\textwidth}{!}{%
\includegraphics{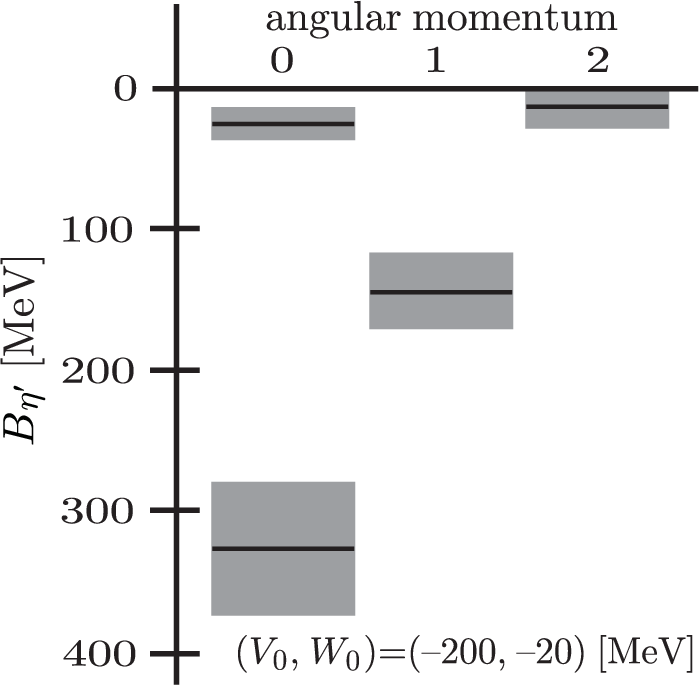}
}
\end{center}
\caption{Level structure of the $\eta'$--$\alpha$ bound states with the potential parameter $(V_0, W_0)=(-200, -20)\ \rm{MeV}$ case. The level width for each state is indicated by the hatched area.}
\label{fig:levelstructure}
\end{figure}

\begin{figure}[htbp]
\begin{center}
\resizebox{0.45\textwidth}{!}{%
\includegraphics{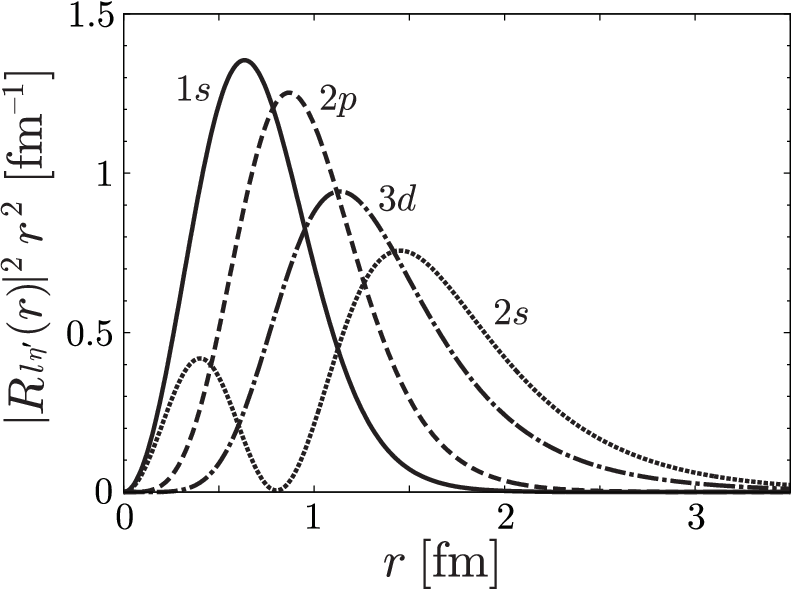}
}
\end{center}
\caption{Calculated density distributions of the $\eta'$ bound state in $\alpha$ are plotted as a function of the radial coordinate $r$ for the potential parameter $(V_0, W_0)=(-200, -20)\ \rm{MeV}$ case.}
\label{fig:fig6-8}
\end{figure}

The calculated binding energies and widths are depicted in Fig. \ref{fig:levelstructure} for $(V_0, W_0)=(-200, -20)\ \rm{MeV}$ case, where the level widths are shown by the hatched area. In this potential case, we have found the four well-separated bound levels. In Fig. \ref{fig:fig6-8}, the radial density distributions of the bound $\eta'$ are shown for the same potential case. 

\begin{figure}[htbp]
\begin{center}
\resizebox{0.48\textwidth}{!}{%
\includegraphics{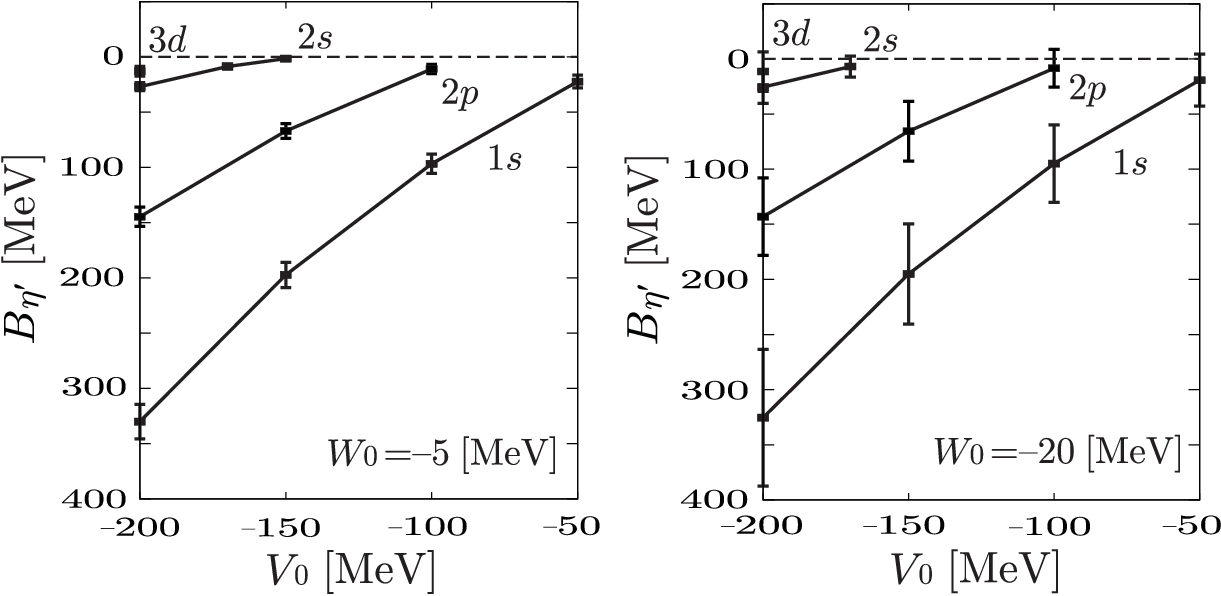} 
}
\end{center}
\caption{Calculated binding energies of the $\eta'$--$\alpha$ bound states are plotted as functions of the potential parameter $V_0$ for two different absorptive potential strength $W_0=-5$ (left) and $-20$ (right) $\rm{MeV}$. The widths and the quantum numbers of bound states are indicated in the figure.}
\label{fig:V0-BE-5_20}
\end{figure}

The calculated results of the $\eta'$--$\alpha$ bound states for various potential depths are shown in Fig. \ref{fig:V0-BE-5_20}. We can see that the binding energies of the bound states are mainly determined by the strength of the attractive potential, while the widths are by the imaginary potential as naturally expected. The calculated values of the binding energies and widths are compiled in Table \ref{table:BE-gamma} for all cases considered here. Here, the binding energies assumed for the calculation of the momentum transfer in Fig. \ref{fig:q} are the $1s$ state with $(V_0, W_0)=(-100, -5)\ \rm{MeV}$ for $B_{\eta'}=96.8\ \rm{MeV}$, and the average of the $1s$ state with $(V_0, W_0)=(-50, -5)\ \rm{MeV}$ and the $2s$ state with $(V_0, W_0)=(-200, -5)\ \rm{MeV}$ for $B_{\eta'}=24.7\ \rm{MeV}$.

\begin{table}[htbp]
\begin{center}
\caption{Calculated binding energies and widths of the $\eta'$--$\alpha$ bound systems are shown in unit of MeV for all cases studied in this article.}
\label{table:BE-gamma}
\begin{tabular}{|c|c|rr|rr|} \hline 
\multirow{3}{*}{$V_0$\,[MeV]} & \multirow{3}{*}{state} & \multicolumn{4}{|c|}{$W_0$\,[MeV]}                                                                                 \\ \cline{3-6}
                                        &                              & \multicolumn{2}{|c|}{-5}                               & \multicolumn{2}{|c|}{-20}                              \\ \cline{3-6}
                                        &                              &  \ $B_{\eta'}$ \  &  \ $\Gamma_{\eta'}$ \     & \  $B_{\eta'}$ \  & \  $\Gamma_{\eta'}$ \      \\ \hline
    -50                               & $1s$                       &  \ 22.4 \           &  \ 11.6 \                        & \ 19.3 \            & \ 47.3 \                          \\ \hline
\multirow{2}{*}{-100}             & $1s$                       &  \ 96.8 \           & \ 17.6 \                         & \ 95.0 \            & \ 70.4 \                          \\
                                        & $2p$                       &  \ 11.1 \           & \ 8.4 \                           & \ 8.5 \             & \ 34.4 \                          \\ \hline
                                        & $1s$                       &  \ 197.3 \          & \ 22.8 \                         & \ 195.0 \          & \ 91.0 \                          \\    
    -150                             & $2s$                        &  \ 1.7 \             & \ 2.4 \                          &                        &                                     \\
                                       & $2p$                        &  \ 67.1 \            & \ 13.5 \                        & \ 65.6 \            & \ 54.1 \                          \\ \hline
\multirow{4}{*}{-200}            & $1s$                        &  \ 330.0 \          & \ 31.3 \                         & \ 325.4 \          & \ 124.0 \                        \\
                                       & $2s$                        &  \ 27.0 \            & \ 7.3 \                          & \ 25.6 \            & \ 29.3 \                          \\
                                       & $2p$                        &  \ 144.7 \          & \ 17.6 \                         & \ 143.1 \          & \ 70.2 \                          \\
                                       & $3d$                        &  \ 13.0 \            & \ 8.9 \                          & \ 11.5 \            & \ 36.0 \                          \\ \hline       
\end{tabular}
\end{center}
\end{table}

\subsection{Formation of the $\eta'$--$\alpha$ bound states in $^6\rm{Li}(\gamma,\it{d})$}
\begin{figure}[htbp]
\begin{center}
\resizebox{0.45\textwidth}{!}{%
\includegraphics{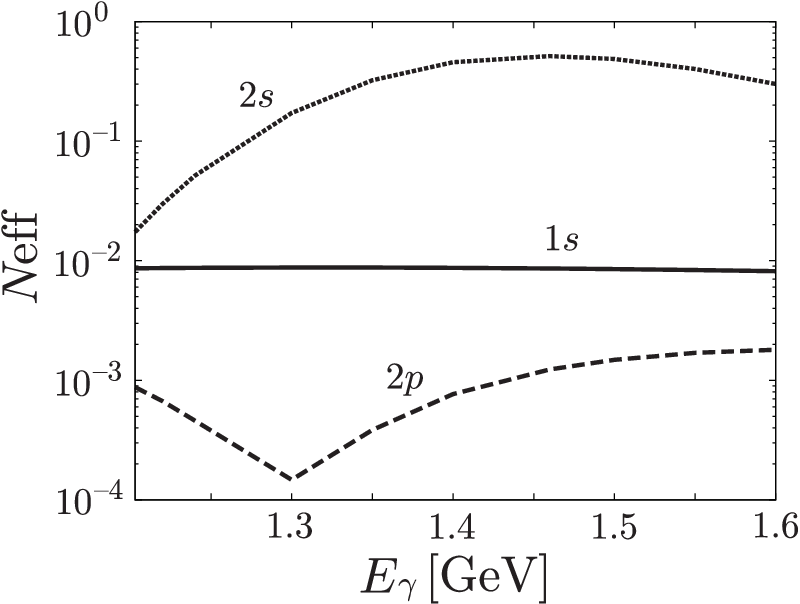}
}
\end{center}
\caption{Photon energy dependence of the calculated effective numbers with $(V_0, W_0)=(-150, -5)\ \rm{MeV}$. The quantum numbers of the $\eta'$ bound states are indicated in the figure.}
\label{fig:Neff-E1}
\end{figure}

We calculate the formation spectra defined by Eq. (\ref{eq:CS}) and show the numerical results in this section. First, we show in Fig. \ref{fig:Neff-E1} the effective numbers for the $1s, 2p, 2s$ bound states calculated with $(V_0, W_0)=(-150, -5)\ \rm{MeV}$ as an example. For each state, the recoilless condition is satisfied at $E_\gamma\sim0.98\ \rm{GeV}$ for the $1s$ state, $E_\gamma\sim1.46\ \rm{GeV}$ for the $2s$ state, and $E_\gamma\sim1.30\ \rm{GeV}$ for the $2p$ state, respectively. We can see from the figure that the effective number for the $2s$ state formation takes the maximum value at the recoilless energy. This behavior of the effective number has been expected since the $d$--$\alpha$ relative motion in the target $^6\rm{Li}$ is described by the $2s$ wave function which has the largest overlap with the $2s$ bound state wave function of the $\eta'$--$\alpha$ system at the recoilless kinematics. In contrast the effective number for the $2p$ state takes the smallest value at the recoilless kinematics. This behavior can be understood as follows. Since the relative wave functions between $\eta'$--$\alpha$ and $d$--$\alpha$ have the similar spatial dimension, they satisfy the approximate orthogonal condition for the states with different quantum numbers. Thus, at the recoilless kinematics, the effective number for the formation of the $2p$ state of $\eta'$--$\alpha$ is strongly suppressed because of the $2s$ wave function for the $d$--$\alpha$ state in the target $^6\rm{Li}$. Finally, the effective number for the $1s$ state does not show the clear $E_\gamma$ dependence, namely the momentum transfer dependence, in this energy region because of the compact wave function of the deeply bound states. If we look more closely the line of the $1s$ state effective number in Fig. \ref{fig:Neff-E1}, we can find that the effective number takes the largest value at $E_\gamma\sim1.3\ \rm{GeV}$ and is slightly suppressed at the recoilless kinematics. This suppression can be understood as in the case of the $2p$ state formation as explained above by considered the approximate orthogonality between $1s$ wave of $\eta'$--$\alpha$ and $2s$ wave of $d$--$\alpha$.

\begin{figure}[htbp]
\begin{center}
\resizebox{0.48\textwidth}{!}{%
\includegraphics{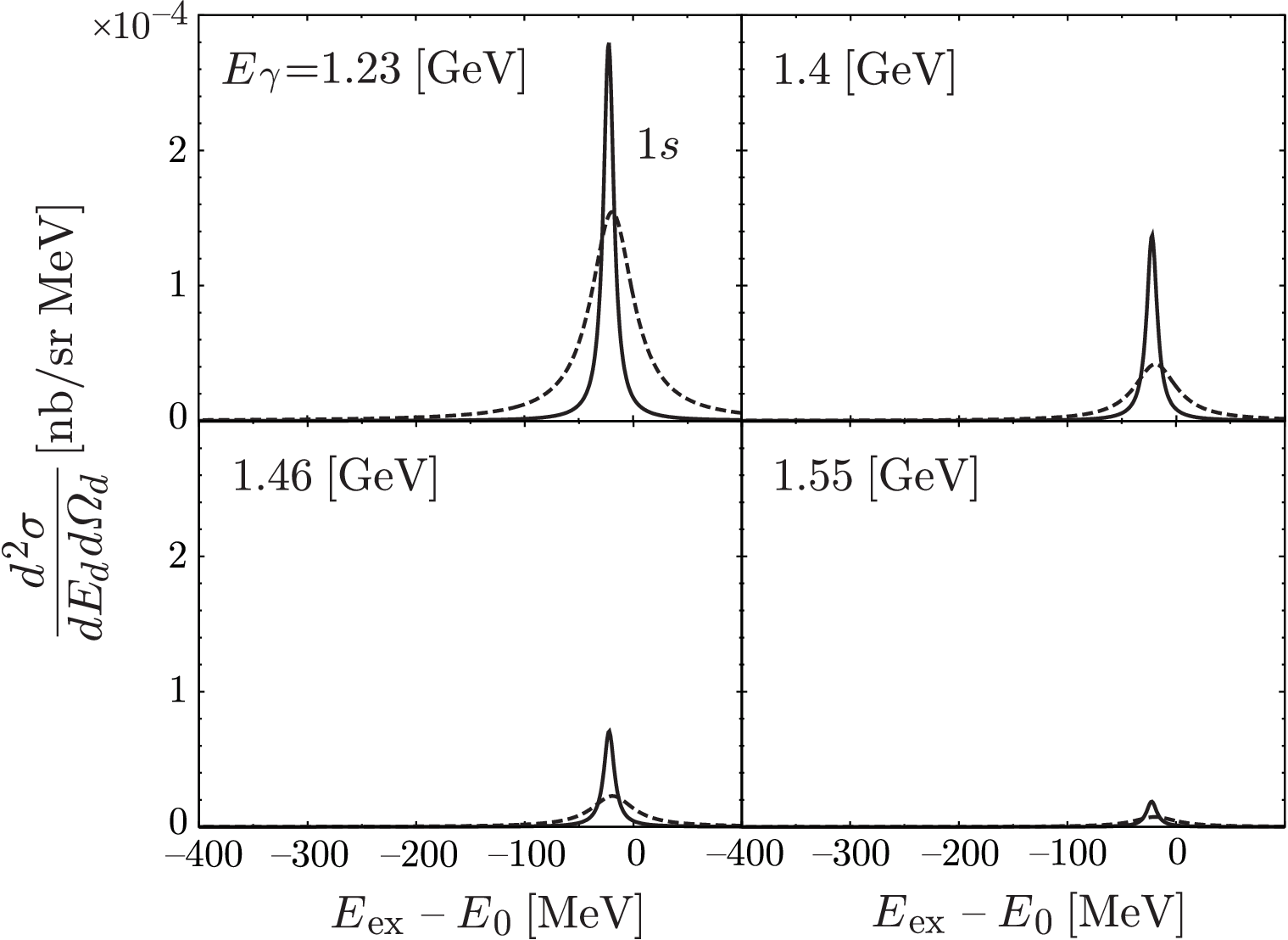}
}
\end{center}
\caption{Expected spectra of the forward $^6\rm{Li}(\gamma,\it{d})$ reaction for the formation of the $\eta'$ bound state in $\alpha$ are plotted as functions of the excitation energy of the $\eta'$ mesic nucleus for four incident photon energies as indicated in the figure. The parameter of the real part of the optical potential is assumed to be $V_0=-50\ \rm{MeV}$ and that of the imaginary part to be $W_0=-5\ \rm{MeV}$ for the solid lines and $-20\ \rm{MeV}$ for the dashed lines. The contribution from the quasi-free $\eta'$ production is not included in these spectra.}
\label{fig:CS-V0_50}
\end{figure}

\begin{figure}[htbp]
\begin{center}
\resizebox{0.48\textwidth}{!}{%
\includegraphics{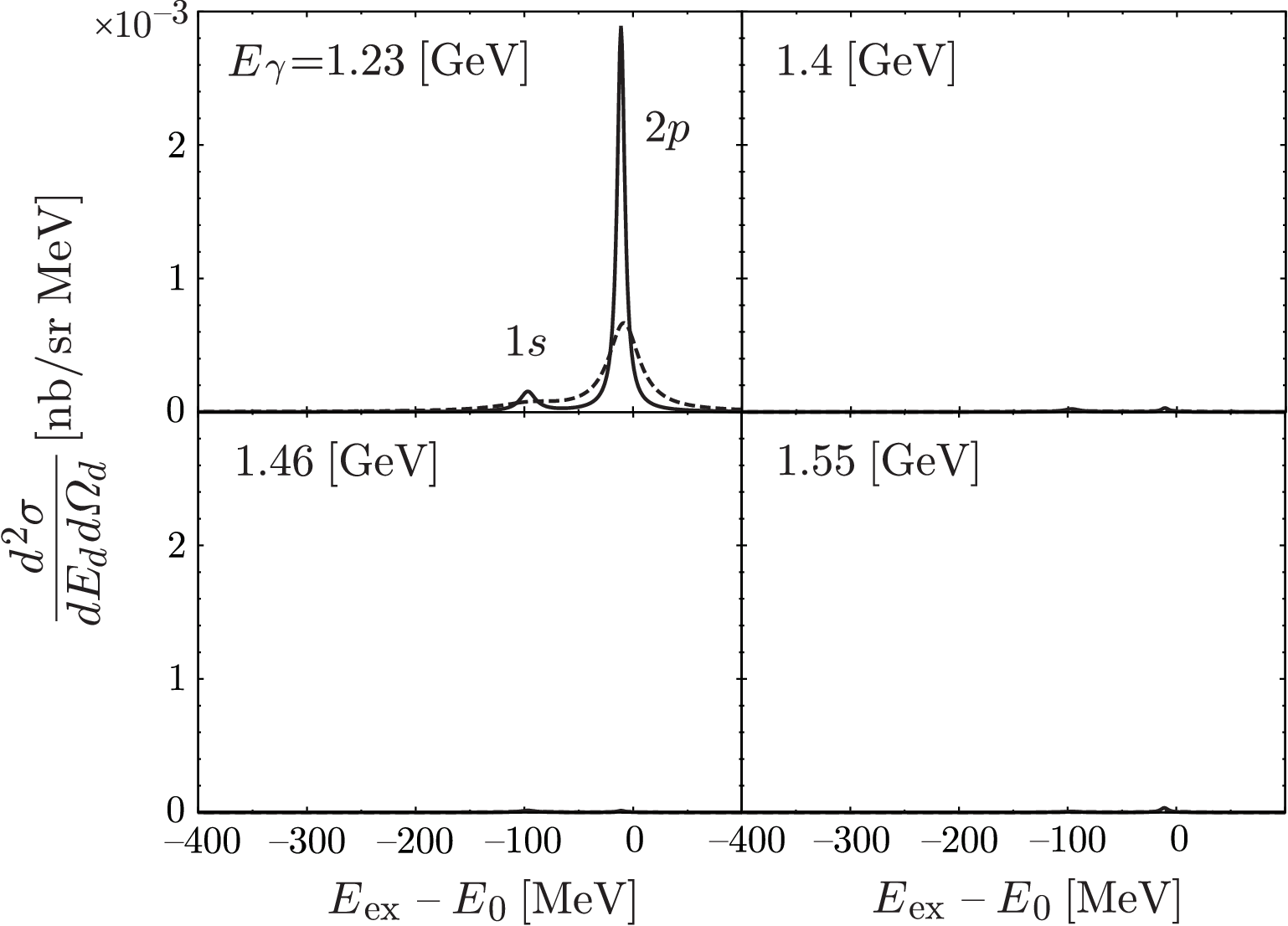}
}
\end{center}
\caption{Same as Fig. \ref{fig:CS-V0_50} except for $V_0=-100\ \rm{MeV}$.}
\label{fig:CS-V0_100}
\end{figure}

\begin{figure}[htbp]
\begin{center}
\resizebox{0.48\textwidth}{!}{%
\includegraphics{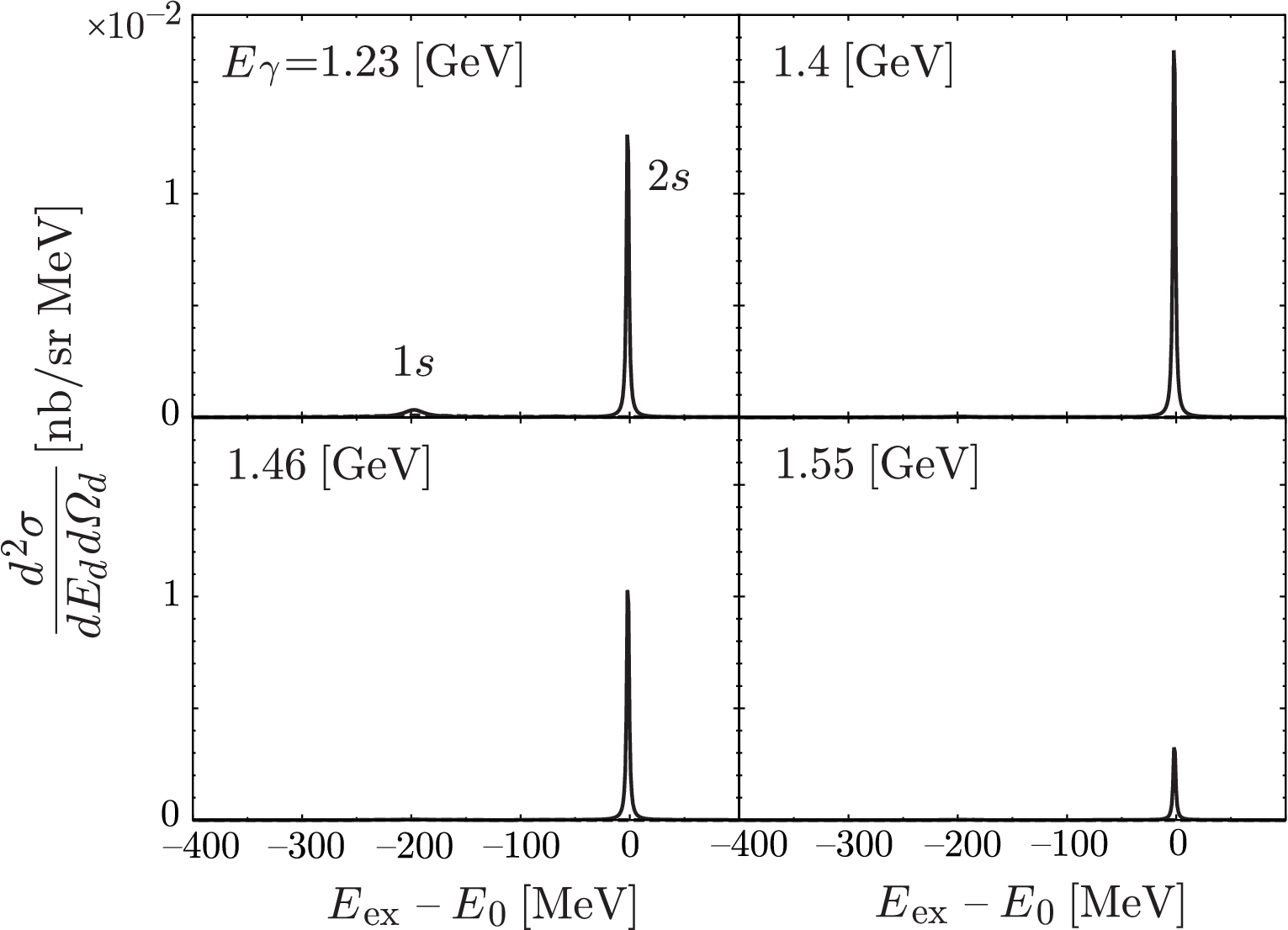}
}
\end{center}
\caption{Same as Fig. \ref{fig:CS-V0_50} except for $V_0=-150\ \rm{MeV}$.}
\label{fig:CS-V0_150}
\end{figure}

\begin{figure}[htbp]
\begin{center}
\resizebox{0.48\textwidth}{!}{%
\includegraphics{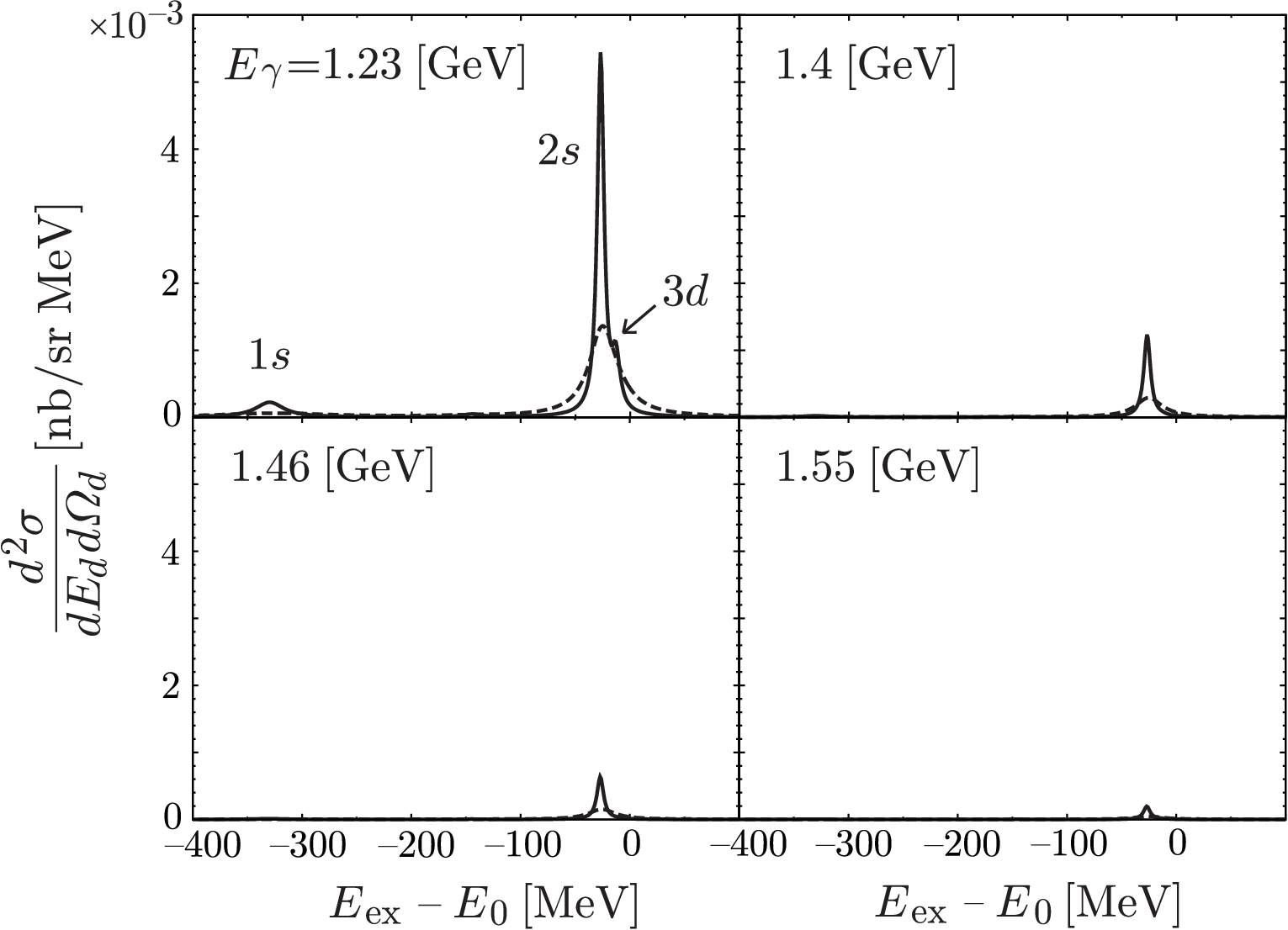}
}
\end{center}
\caption{Same as Fig. \ref{fig:CS-V0_50} except for $V_0=-200\ \rm{MeV}$.}
\label{fig:CS-V0_200}
\end{figure}

We show below the calculated formation spectra of the $\eta'$--$\alpha$ bound states. First, we show the spectra for the $V_0=-50\ \rm{MeV}$ case in Fig. \ref{fig:CS-V0_50}, which is the weakest attractive potential case considered in this article, for four incident photon energies. The potential parameter of the absorptive part is assumed to be $W_0=-5\ \rm{MeV}$ and $-20\ \rm{MeV}$. The spectra are plotted as the functions of the excitation energy $E_{\rm{ex}}$ of the $\eta'$ mesic nucleus, which are defined as,

\begin{eqnarray*}
E_{\rm{ex}}-E_0=p_\gamma -T_d-m_{\eta'}-S_d, 
\end{eqnarray*}

\noindent where $E_0$ indicates the threshold energy of the $\eta'$ production in the $^6\rm{Li}(\gamma,\it{d})$ reaction. All spectra calculated in this article are plotted for the same range of the excitation energy in Figs. \ref{fig:CS-V0_50}$-$\ref{fig:CS-V0_200} and \ref{fig:CS-dsh} (b). The peaks in $E_{\rm{ex}}-E_0<0$ region mean the formation of the bound states. We found that we can see the peak structure in the spectra for the formation of the $1s$ state. The absolute value of the cross section is as small as the order of around $0.1\ \rm{pb/sr/MeV}$ in this case. We also found that the cross section become smaller for the higher incident energies.

We also show the calculated spectra for other cases with different potential depths in Figs. \ref{fig:CS-V0_100}$-$\ref{fig:CS-V0_200}. In Fig. \ref{fig:CS-V0_100}, we can see the peak due to the formation of the $1s$ state of $\eta'$--$\alpha$ around $E_{\rm{ex}}-E_0\sim-100\ \rm{MeV}$ for $E_\gamma=1.23\ \rm{GeV}$, and the peak due to the formation of $2p$ bound states is also seen largely at $E_{\rm{ex}}-E_0\sim-10\ \rm{MeV}$. The cross section of the $2p$ bound state formation is large at $E_\gamma=1.23\ \rm{GeV}$ because the matching condition between the momentum and the angular momentum transfer is satisfied at this energy. Since the binding energy is small for the $2p$ state for $V_0=-100\ \rm{MeV}$, the momentum transfer is around $200\ \rm{MeV/c}$ at $E_\gamma=1.23\ \rm{GeV}$ as shown in Fig. \ref{fig:q} by the solid line for the $B_{\eta'}=0$ case. If we roughly estimate the radius $R$ of the $\alpha$ particle to be around $1\ \rm{fm}$, the reaction with the angular momentum transfer $\Delta l=1$ satisfies the matching condition $\Delta l \sim Rq$. Thus, the formation of $2p$ bound state of $\eta'$--$\alpha$ from the $2s$ bound state of $d$--$\alpha$ (=$^6\mathrm{Li}$) is enhanced at this energy for the $V_0=-100\ \rm{MeV}$. In Fig. \ref{fig:CS-V0_150}, we can see the highest peak in all cases considered in this article in the spectra for $E_\gamma=1.4\ \rm{GeV}$. In this case with $V_0=-150\ \rm{MeV}$, there is a $2s$ bound state close to the threshold. And this state can be produced in the recoilless condition around $E_\gamma=1.4\ \rm{GeV}$. Since the relative wave function of $d$--$\alpha$ in $^6\rm{Li}$ is considered to be $2s$ state, the $2s$ bound state of the $\eta'$--$\alpha$ system is largely produced as a substitutional state in this case. And as expected from the effective numbers shown in Fig. \ref{fig:Neff-E1}, the peak of the $\eta'$ bound state formation in the $2s$ level is the highest one in the spectra for all cases shown in Fig. \ref{fig:CS-V0_150}. We can see in Fig. \ref{fig:CS-V0_200} the complex peak structure composed of the $2s$ and $3d$ bound states around the threshold at $E_\gamma=1.23\ \rm{GeV}$. We also see the deepest bound state contribution at $E_{\rm{ex}}-E_0\sim-330\ \rm{MeV}$ in this case.
 
In all cases considered here, we find that the formation spectra have the strong tendency to be smaller for the higher photon energies, while the relative strength of each peak can be reasonably interpreted by the behavior of the effective numbers. The photon energy dependence is naturally understood by the reduction of the elementary cross section shown in Fig. \ref{fig:etaprime-eleCS-0}, which is due to the fragile nature of the deuteron expressed by the form factor in our formulation. As for the behavior of the effective numbers, the energy dependence of the shallow $\eta'$ bound states such as $2s$ and $2p$ states is well understood by the matching condition between the momentum and the angular momentum transfer of the reaction. The energy dependence of the formation of the deepest $1s$ bound state of $\eta'$ is found to be weak because of the compactness of the wave function. And the peak height of the $1s$ state formation is relatively small for all cases because of the small overlap with the $2s$ radial wave function of the $d$--$\alpha$ initial state and the larger width due to the $\eta'$ nuclear absorption than shallower $\eta'$ bound states. Thus, to obtain the larger formation cross section of the $\eta'$ bound states it is better to consider to use the lower photon energies for the $\eta'$ production and to choose the photon energy and/or $\eta'$ state to satisfy the matching condition of the momentum and the angular momentum transfer. 

\subsection{Effects of the possible shrinkage of the quasi-deuteron in $^6\rm{Li}$ target}
In this section, we consider the effects of the possible shrinkage of the quasi-deuteron in $^6\rm{Li}$ target and the possible enhancement of the elementary cross section. 

The S- and D-wave deuteron wave functions $u(r)$ and $w(r)$ shown in section \ref{sec:Effective number formalism} are parametrized in Ref. \cite{Bonn} as,

\begin{align}
\begin{split}
u(r)&=\sum_j C_j \exp \left(-m_j r\right),  \\
w(r)&=\sum_j D_j \exp \left(-m_j r\right)\left(1+\frac{3}{m_j r}+\frac{3}{\left(m_j r\right)^2}\right),
\label{eq:parametrization}
\end{split}
\end{align}

\noindent where $C_j$'s, $D_j$'s, and $m_j$'s are the constants given in Ref. \cite{Bonn}. To simulate the shrinkage, we introduce the scale factor $\alpha$ ($0<\alpha<1$) and vary the parameters $m_j$, $C_j$, and $D_j$ in Eq. (\ref{eq:parametrization}) as,

\begin{align}
\begin{split}
m_j &\rightarrow \frac 1\alpha m_j, \\
C_j &\rightarrow \frac 1{\sqrt{\alpha}} C_j, \\
D_j &\rightarrow \frac 1{\sqrt{\alpha}} D_j. 
\label{eq:dshrink-scale}
\end{split}
\end{align}

\noindent This scaling clearly does not change the normalization condition,

\begin{eqnarray*}
\int^\infty_0 dr \left[\left(u(r)\right)^2+\left(w(r)\right)^2\right]=1,
\end{eqnarray*}

\noindent while the root-mean-square radius for the deuteron scales as, 

\begin{equation*}
\sqrt{\left<r_d^2\right>} \rightarrow \alpha \sqrt{\left<r_d^2\right>}.
\end{equation*}

\begin{figure}[htbp]
\begin{center}
\resizebox{0.47\textwidth}{!}{%
\includegraphics{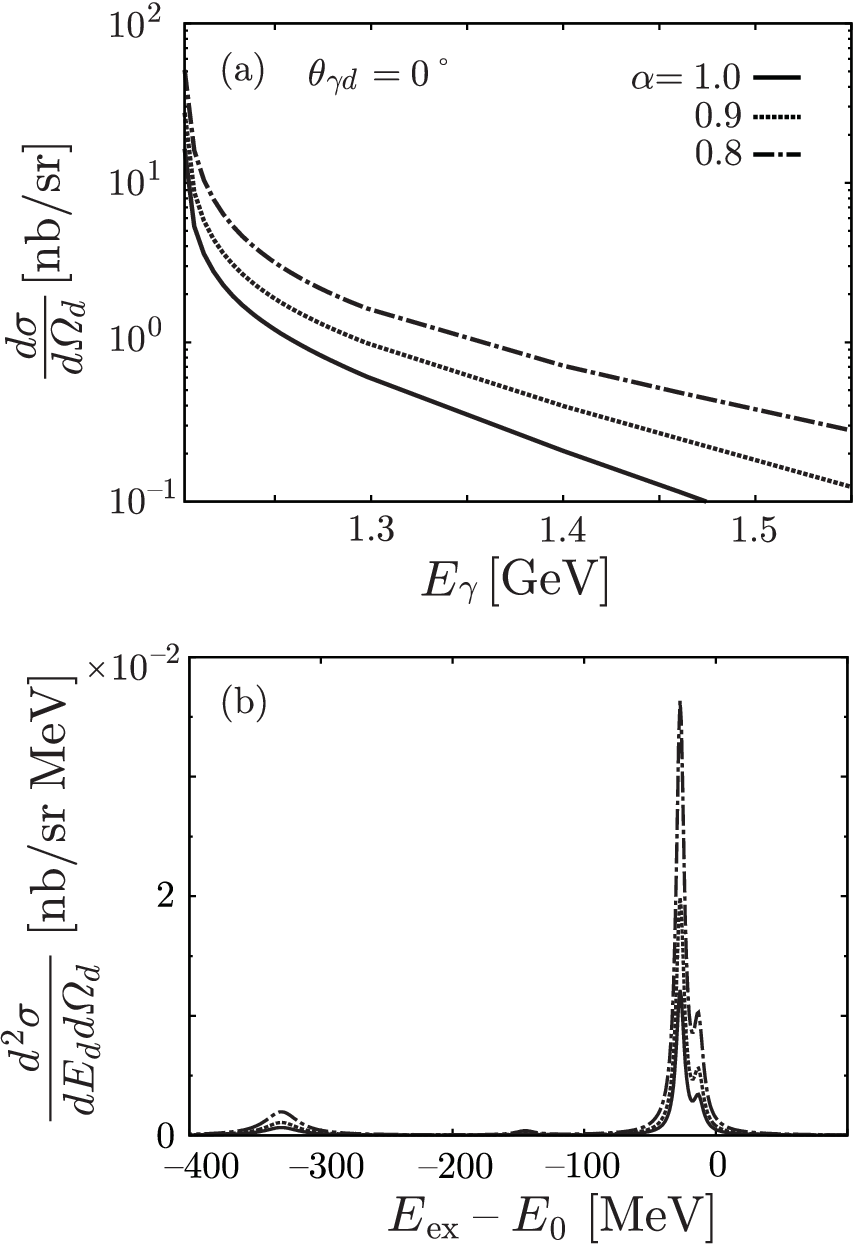}
}
\end{center}
\caption{(a) Incident photon energy dependence of the differential cross section of the elementary $\gamma+d \rightarrow d+\eta'$ reaction at $\theta_{\gamma d}=0^\circ$ in the laboratory frame with $c=0.1\ \rm{fm}$. The values of the scale factor $\alpha$ introduced in Eq. (\ref{eq:dshrink-scale}) are indicated in the figure. (b) Calculated spectra of the forward $^6\rm{Li}(\gamma,\it{d})$ reaction with the deuteron shrink effects for the formation of the $\eta'$ bound state in the $\alpha$ particle are plotted as functions of the excitation energy of the $\eta'$ mesic nucleus for the incident photon energy $E_\gamma=1.208\ \rm{GeV}$ and the potential parameters $(V_0, W_0)=(-200, -5)\ \rm{MeV}$ case. The contribution from the quasi-free $\eta'$ production is not included in this spectra. The values of the scale factor $\alpha$ are the same as indicated in the figure (a).} 
\label{fig:CS-dsh}
\end{figure}

Then we show the elementary cross section and the formation spectra of the $\eta'$--$\alpha$ system for the different $\alpha$ values to know the role of the deuteron form factor and the effects of the deuteron shrinkage in Fig. \ref{fig:CS-dsh}. We can see from the Fig. \ref{fig:CS-dsh} (a) that the value of the differential cross section at $\theta_{\gamma d}=0^\circ$ of the elementary $\gamma+d \rightarrow d+\eta'$ reaction become larger by the deuteron shrinkage because of the larger deuteron form factor at the high momentum transfer for large photon energies. Accordingly, we have the larger formation spectra of the $\eta'$--$\alpha$ system as shown in Fig. \ref{fig:CS-dsh} (b). We can see from the Fig. \ref{fig:CS-dsh} that the shrinkage of the deuteron enhances the cross section as we have expected and the size of the enhancement is around factor 3 for the scale factor $\alpha=0.8$ case with the incident photon energy $E_\gamma=1.208\ \rm{GeV}$ and the potential parameters $(V_0, W_0)=(-200, -5)\ \rm{MeV}$. 

\section{Conclusions} \label{sec:Conclusions}
We have reported the first calculated results of the formation cross section of the $\eta'(958)$ mesic nucleus in $\alpha$ by the $^6\rm{Li}(\gamma,\it{d})$ reaction. We have improved the theoretical formula reported in Ref. \cite{Ikeno_gamma_d} in the following four points, (i) the inclusion of the distortion effects for emitted deuteron, (ii) the evaluation of the elementary cross section, (iii) the usage of the realistic $\alpha$ density, and (iv) the inclusion of the center-of-mass correction of the reaction. We have shown the numerical results for the various cases with the different $\eta'$--$\alpha$ interaction strengths and incident photon energies.

We have found that the relative strength of each subcomponent of formation spectra is reasonably understood in the context of the matching condition as we expected. On the other hand, we have also found that the absolute strength of the whole spectrum has strong tendency to be smaller for higher incident photon energies. This tendency can be naturally understood by the fragile deuteron form factor. We have checked the effects of the form factor by considering the possible shrinkage of the deuteron in the $^6\rm{Li}$ target. As the conclusion of this article, the order of magnitude of the formation spectra of $\eta'$ mesic nucleus in $\alpha$ particle in the $^6\rm{Li}(\gamma,\it{d})$ reaction is $10\ \rm{pb/sr/MeV}$ or less as shown in Figs. \ref{fig:CS-V0_50}$-$\ref{fig:CS-V0_200} and the formation cross section tends to be larger for the smaller incident photon energies because of the smaller momentum transfer to the deuteron. We hope that our results stimulate the new experimental studies of the $\eta'$ mesic nucleus, even though it seems difficult to observe these states.

\section*{Acknowledgements}
We acknowledge the fruitful discussions with H. Fujioka and T. Ishikawa from the beginning of this research. We would like to thank E. Hiyama for providing us the realistic density of the $\alpha$ particle. This work is partly supported by the Grants-in-Aid for Scientific Research No. 24540274 and 16K05355 (S.H.), No. 26400275 (H.N.), and No. 15H06413 (N.I.) in Japan.


\begin{thebibliography}{}

\bibitem{Hatsuda1994}
T. Hatsuda and T. Kunihiro, Phys. Rep. \textbf{247}, (1994) 221-367, \doi{10.1016/0370-1573(94)90022-1}.


\bibitem{Yamazaki2012}
T. Yamazaki, S. Hirenzaki, R. S. Hayano, and H. Toki, Phys. Rep. \textbf{514}, (2012) 1-87, \doi{10.1016/j.physrep.2012.01.003}.


\bibitem{Hirenzaki1991}
S. Hirenzaki, H. Toki, and T. Yamazaki, Phys. Rev. \textbf{C44}, (1991) 2472-2479, \doi{10.1103/PhysRevC.44.2472}.


\bibitem{Gilg}
H. Gilg {\it et al.}, Phys. Rev. \textbf{C62}, (2000) 025201, \doi{10.1103/PhysRevC.62.025201}.


\bibitem{Itahashi2000}
K. Itahashi {\it et al.}, Phys. Rev. \textbf{C62}, (2000) 025202, \doi{10.1103/PhysRevC.62.025202}.


\bibitem{Ikeno_gamma_d}
N. Ikeno, J. Yamagata-Sekihara, H. Nagahiro, D. Jido, and S. Hirenzaki, Phys. Rev. \textbf{C84}, (2011) 054609, \doi{10.1103/PhysRevC.84.054609}.


\bibitem{Nagahiro2005}
H. Nagahiro and S. Hirenzaki, Phys. Rev. Lett. \textbf{94}, (2005) 232503, \doi{10.1103/PhysRevLett.94.232503}.


\bibitem{Nagahiro2013}
H. Nagahiro, D. Jido, H. Fujioka, K. Itahashi, and S. Hirenzaki, Phys. Rev. \textbf{C87}, (2013) 045201, \doi{10.1103/PhysRevC.87.045201}.


\bibitem{Itahashi2012}
K. Itahashi {\it et al.}, Prog. Theor. Phys. \textbf{128}, (2012) 601-613, \doi{10.1143/PTP.128.601}.


\bibitem{Kelkar2013}
N. G. Kelkar, K. P. Khemchandani, N. J. Upadhyay, and B. K. Jain, Rep. Prog. Phys. \textbf{76}, (2013) 066301, \doi{10.1088/0034-4885/76/6/066301}.


\bibitem{Krusche2014}
B. Krusche and C. Wilkin, Prog. Part. Nucl. Phys. \textbf{80}, (2014) 43-95, \doi{10.1016/j.ppnp.2014.10.001}.


\bibitem{Machner2015}
H. Machner, J. Phys. \textbf{G42}, (2015) 043001, \doi{10.1088/0954-3899/42/4/043001}.


\bibitem{Haider2015}
Q. Haider and L. C. Liu, Int. J. Mod. Phys. \textbf{E24}, (2015) 1530009, \doi{10.1142/S021830131530009X}.


\bibitem{Ent_0.73}
R. Ent, H. P. Block, J. F. A. van Hienen, G. van der Steenhoven, J. F. J. van den Brand {\it et al.}, Phys. Rev. Lett. \textbf{57}, (1986) 2367-2370, \doi{10.1103/PhysRevLett.57.2367}.


\bibitem{totCS_etaprime}
I. Jaegle {\it et al.}, Eur. Phys. J. \textbf{A47}, (2011) 11, \doi{10.1140/epja/i2011-11011-x}.


\bibitem{Fellinger}
M. Fellinger, E. Gutman, R. C. Lamb, F. C. Peterson, L. S. Schroeder, R. C. Chase, E. Coleman, and T. G. Rhoades, Phys. Rev. Lett. \textbf{22}, (1969) 1265-1269, \doi{10.1103/PhysRevLett.22.1265}.


\bibitem{Bonn}
R. Machleidt, K. Holinde, and Ch. Elster, Phys. Rep. \textbf{149}, (1987) 1-89, \doi{10.1016/S0370-1573(87)80002-9}.


\bibitem{PDG}
Particle Data Group, Review of Particle Physics (RPP), Phys. Rev. \textbf{D86}, (2012) 010001, \doi{10.1103/PhysRevD.86.010001}.


\bibitem{Koike}
T. Koike and T. Harada, Nucl. Phys. \textbf{A804}, (2008) 231-273, \doi{10.1016/j.nuclphysa.2008.01.015}.


\bibitem{Nagahiro2006}
H. Nagahiro, M. Takizawa, and S. Hirenzaki, Phys. Rev. \textbf{C74}, (2006) 045203, \doi{10.1103/PhysRevC.74.045203}. 


\bibitem{Jido2012}
D. Jido, H. Nagahiro, and S. Hirenzaki, Phys. Rev. \textbf{C85}, (2012) 032201, \doi{10.1103/PhysRevC.85.032201}. 


\bibitem{Nagahiro2012}
H. Nagahiro, S. Hirenzaki, E. Oset, and A. Ramos, Phys. Lett. \textbf{B709}, (2012) 87-92, \doi{10.1016/j.physletb.2012.01.061}. 


\bibitem{Czerwinski2014}
E. Czerwinski {\it et al.}, Phys. Rev. Lett. \textbf{113}, (2014) 062004, \doi{10.1103/PhysRevLett.113.062004}.


\bibitem{Nanova2012} 
M. Nanova {\it et al.} [CBELSA/TAPS Collaboration], Phys. Lett. \textbf{B710}, (2012) 600-606, \doi{10.1016/j.physletb.2012.03.039}.

\bibitem{Nanova2013} 
M. Nanova {\it et al.} [CBELSA/TAPS Collaboration], Phys. Lett. \textbf{B727}, (2013) 417-423, \doi{10.1016/j.physletb.2013.10.062}.


\bibitem{Bass2014}
S. D. Bass and A. W. Thomas, Acta Phys. Polon. \textbf{B45}, (2014) 627, \doi{10.5506/APhysPolB.45.627}. 


\bibitem{Hiyama_private}
E. Hiyama, private communication, (2015).


\bibitem{Hiyama}

E. Hiyama, B. F. Gibson, and M. Kamimura, Phys. Rev. \textbf{C70}, (2004) 031001, \doi{10.1103/PhysRevC.70.031001}.


\bibitem{eleCS_eta}
J. Weiss {\it et al.}, Eur. Phys. J. \textbf{A11}, (2001) 371-374, \doi{10.1007/s100500170047}.


\bibitem{eleCS_eta_data}
P. Hoffmann-Rothe {\it et al.}, Phys. Rev. Lett. \textbf{78}, (1997) 4697-4700, \doi{10.1103/PhysRevLett.78.4697}.

\end{thebibliography}
\end{document}